\documentstyle[preprint,aps,epsf]{revtex}

\newcommand{\beq}{\begin{eqnarray}}
\newcommand{\eeq}{\end{eqnarray}}

\newcommand{\btem}{\bibitem}

\begin{document}


\draft

\title{Pad\'{e} Improvement of the Free Energy \\ in High Temperature
 QCD}

\author{T. Hatsuda}

\address{Institute of Physics, University of Tsukuba,
 Tsukuba, Ibaraki 305, Japan}

 
\maketitle

\begin{abstract}
   Pad\'{e} approximants (PA's) are constructed from the 
 perturbative coefficients  of the free energy through $O(g^5)$ 
 in hot QCD. 
 Pad\'{e} summation is shown to reduce the  renormalization-scale
  dependence substantially even at temperature ($T$) as low as  $250$ MeV.
 Also, PA's  predict that the free energy does not deviate
 more than 10 \% from the Stefan-Boltzmann limit for
 $T > 250$ MeV.

\end{abstract}


\narrowtext

 Properties of quantum chromodynamics (QCD)  at high temperature ($T$)
  acquire a lot of attention in relation to the physics of the early
 universe and of the relativistic heavy ion collisions \cite{review}.
 In non-abelian gauge theories, 
 naive expectation, based on the asymptotic freedom,
  is that the perturbation theory works
  as far as the temperature is high enough  \cite{asy}.
   However, this may not be the case because of
  the infrared sensitivity in the high-order terms beyond
 $O(g^6)$ for the free energy with $g$ being the QCD coupling
 constant \cite{linde}.

 Another practical problem is the asymptotic nature of the
 perturbation series, which can be seen from the expansion
  recently established  through $O(g^5)$  \cite{g5}.
 If one takes the expansion literally, 
 the perturbation theory works only for $T > $  2 GeV which
 is order of magnitude larger than the critical temperature ($T_c$)
 of the QCD phase transition. 
  On the other hand, the lattice QCD simulations for 
 energy density, pressure and entropy density show that 
 they approach  the Stefan-Boltzmann (SB) limit 
  rather quickly above 
 $T_c \simeq 150$ MeV  \cite{ah}.

 This poses a question that whether one could reconcile the 
 non-perturbative lattice result with the perturbative expansion
in the region  $ T_c < T < 2 $ GeV  by making
  proper resummation of the perturbation series.
 From the experimental point of view,
   the highest  temperature which one could access in the
 relativistic heavy-ion colliders are at most 300-500 MeV.
 Therefore,  most of the future data reflect the physics
 in the above temperature interval.

 If the perturbative expansion has an asymptotic nature,
  the truncated series
 shows a large fluctuation 
 for medium/strong couplings  by the change of the order of the 
 truncation.  
  Nevertheless, some information on the
 full result is reflected in  the behavior of the fluctuation and
  the coefficients
 of the perturbative expansion.
  This is the place where
 resummation method such as the Pad\'{e} approach
 could play a role \cite{pade}.
 In fact, the Pad\'{e} summation has been successfully applied to the
 perturbative QCD series in high energy processes:
 it `postdicts' the known higher-order terms, and
 also removes the renormalization-scheme dependence \cite{ek,gardi}.

In the following, we will apply the Pad\'{e} approach for free energy
 of QCD at finite $T$ with vanishing chemical potential and quark masses.
 The general structure of 
 the free energy $F(T,\mu)$ for massless $n_f$-flavors in $\overline{MS}$
  scheme
 through $O(g^5)$ reads \cite{g5,BN}
\beq
\label{RR}
R(\mu) \equiv {F(T,\mu) \over F_{SB}(T)} 
 = 1 + f_2 \ \bar{g}^2(\mu) + 
      f_3 \ \bar{g}^3(\mu) + \\ \nonumber
  f_4 (\mu /T, \ln \bar{g})\  \bar{g}^4(\mu)
  + f_5(\mu /T) \ \bar{g}^5(\mu),
\eeq
where $\mu$ is a renormalization-scale,
 $\bar{g} \equiv g/2\pi$ and $F_{SB}$ is the SB (classical) limit
 of the free energy. 
 Since the explicit form of the coefficients $f_i$ is 
 given in \cite{g5,BN}, we will not recapitulate them here.

 We plot $R(\mu=2 \pi T)$ for $n_f=4$ in Fig.1(a) as a function of 
 $\alpha_s(\mu = 2 \pi T) $ with $\alpha_s \equiv g^2/4\pi$ 
 (see the discussions below for this choice of $\mu$).
  The solid lines in Fig.1(a) have a large fluctuation 
 as one increases 
 the order of the perturbation.
   $\alpha_s(2\pi T)$ used in Fig.1(a) is shown in Fig.1(b),
 where  we take the two-loop $\beta$-function with $n_f =4$ for
 simplicity.
  Instead of  
 expanding $\alpha_s$  by $1/\ln(\mu /\Lambda )$,
 we calculated the running of $\alpha_s$ numerically
  with an initial condition
 $\alpha_s(\mu = 5 {\rm GeV}) = 0.21$ obtained in 
 the $\overline{MS}$ scheme  \cite{PDG}.

 Before applying the Pad\'{e} summation,
 let us first examine the renormalization-scheme dependence of 
 $R(\mu)$. Up to  $O(g^5)$ for the $\beta$-function, the
 scheme dependence is equivalent to the renormalization-scale
 dependence \cite{PMS}.  The solid line with `pert.'  in Fig.2 shows 
 $R(\mu)$ for $n_f =4$  as a function of $\mu$ at $T=250$ MeV.
  This temperature is rather low but is still 1.7 times larger than $T_c$.
 In Fig.2,  $R(\mu)$ shows a sizable $\mu$ dependence and  
 even becomes unstable (negative pressure) for low $\mu$.
  $R(\mu)$ keeps this substantial $\mu$-dependence unless
 $T$ is extremely large.

  The principle of minimum sensitivity (PMS) criterion \cite{PMS}
  does not work in the above situation,  
 since there is no solution of the 
  stability condition $dR(\mu)/d\mu =0$.
 The fastest convergence criterion (FAC) gives an 
 unphysically large  value $\mu = 37.7 \pi T$ (for $n_f=4$) \cite{g5}.
 The criterion motivated by the Brodsky-Lepage-Mackengie (BLM)
 \cite{BLM} suggests $\mu = (0.95 \sim 4.4) \pi T$
 but gives rise to a  series reliable only for 
  $T > 2$ GeV \cite{g5}.
 Thus, any  choice of $\mu$ cannot solve the problem at hand, and we
 really need a summation of the series.

  Perturbation series eq.(\ref{RR}) has two 
 different features from that for high energy processes: 
\begin{description}
\item[(i)]
 Odd powers of $\bar{g}$ appear.
\item[(ii)] 
  There arises $\ln \bar{g}$ in the coefficients.
 In fact,  the coefficient $f_4$ depends linearly on
 $\ln \bar{g}$. Also, it is expected to appear at $O(g^6)$ level \cite{BN}.
\end{description}
  Since the standard Pad\'{e} approximants  are based  on the ratio of  
  polynomials,
 (ii) is a new feature beyond the
 standard method in a strict sense.
   In this paper, we  take a simple procedure that
  $\ln \bar{g}$ is regarded as a part of the coefficients $f_{i}$.
 
Let us write down the general form of the Pad\'{e} approximants (PA's)
\beq
R^{[N/M]} (\mu) 
= { 1 + \sum_{n=1}^N c_n \bar{g}^n 
  \over
 1 + \sum_{m=1}^M d_m \bar{g}^m }.
\eeq
$c_n$ and $d_m$ are the functions of $f_i$.
Since we have 5 coefficients $f_i$ ($1 \le i \le 5 $),
 we can construct PA's satisfying  $N+M \leq 5$.
 For example, the highest PA's are [3/2], [2/3], [4/1] and [1/4].

PA's discussed above are the most ``naive'' Pad\'{e} approximants (NPA's).
 They have some  unnatural feature 
 from the physics point of view:
 NPA's contain terms proportional to $\bar{g}$ both 
 in the numerator and the denominator.
  On the other hand, in eq.(\ref{RR}), we never have $O(\bar{g})$ term
 because of a trivial reason (the free energy has no external legs).
  Although NPA's assure that 
  their Tayler expansion by $\bar{g}$ does not contain
 $O(\bar{g})$-term by the condition $c_1=d_1$, 
   it is still unnatural that fictitious $O(\bar{g})$-term play 
  a  crucial role to determine  higher orders.

  Therefore, one may also try  alternative PA's
 by assuming $c_1 = d_1 =0$. In this 
  ``constrained'' Pad\'{e} Approximants (CPA's), one 
 can have PA's one-step beyond, i.e.   $N+M \leq 6$.
 For example, the highest CPA's are 
  [4/2], [2/4] and [3/3].
 It turns out that 
  the coefficients ($c_i$, $d_i$) in CPA's are much simpler than those of  
 NPA's because of the condition $c_1=d_1=0$.
 In particular, ($c_i$, $d_i$) in NPA's usually become 
 complicated functions of $\ln \bar{g}$, while
 they simple depend linearly on $\ln \bar{g}$ in CPA's.

 Besides NPA's and CPA's, one can also start with 
 the effective charge $S(\mu) \equiv (R(\mu)-1)/f_2 $.
 The corresponding PA's for effective charge (EPA's) read
$S^{[N/M]} (\mu)  = \bar{g}^2 \ ( 1 + \sum_{n=1}^N c_n \bar{g}^n )/
 ( 1 + \sum_{m=1}^M d_m \bar{g}^m ).$
 In this case,  $N+M \leq 3$ holds and
 highest PA's are [2/1] and [1/2].
 However, one can show that [2/1]-EPA and
 [1/2]-EPA are equivalent to [4/1]-NPA and
 [3/2]-NPA respectively, so they are not independent from
 NPA. 
 
We have tried all possible PA's mentioned above and 
the following is a summary.

\begin{description}
\item[(a)] 
 \ \ 
 [3/2]-NPA, [2/3]-NPA, [4/2]-CPA, and [2/4]-CPA give results 
  qualitatively consistent with each other.

  The solid (dashed) lines in Fig.2 show the $\mu$ dependence
 of CPA's (NPA's) at $T=250$ MeV. The renormalization-scale
 dependence is reduced substantially compared to the 
 original series (the solid line with `pert.').
 This  is one of the justifications that  the Pad\'{e}
 approximants point to the right direction.
  Similar $\mu$ independence has been reported on
 PA's for the Bjorken sum rules at high $Q^2$ \cite{ek}.
 In the latter case, the approximate invariance
 of PA's under the Euler transform is a key to the weak 
  scale-dependence \cite{gardi}.  The similar argument is expected to hold 
  at finite $T$.

\item[(b)]
\ \   [4/1]-NPA and [1/4]-NPA
  turn out to develop a pole 
  in  the denominator at  $\alpha_s \sim 0.04$ and $0.06$, respectively.
  [3/3]-CPA also has poles at 
   $\alpha_s \sim 0.12$ and $ \sim 0.3$.
  Whether they have  real physical meaning or
  are the artifact of the approximation is not known.

\item[(c)]
\ \  Low-order PA's, namely $N+M \leq 4$ for NPA's, $N+M \leq 5$ for CPA's 
 and $N+M \leq 2$ for EPA's, do not 
 properly `postdict' the existence of 
 $\ln \bar{g}$ in $f_4$ or the absence of
 $\ln \bar{g}$ in $f_5$. This implies that every known information
 of $f_i  (i \leq 5)$ is required to 
 construct proper PA's.

\end{description}

 Let us now focus on the successful cases in item (a) above.
  Because of the small $\mu$ dependence of PA's in (a),
 the choice of $\mu$ is not a serious problem anymore 
 as far as $\mu > 1 $GeV for $T > 250 $MeV.
 The BLM scheme, which requires the leading $n_f$ dependence
 to vanish, gives $\mu \simeq 4.4 \pi T$ ( $ \mu \simeq \pi T$)  when
 it is applied to the coefficient $f_4$ ( $f_5$).
 Based on these observations together with the fact that
 $2 \pi T$ is the lowest non-zero Matsubara frequency for gluons,
 we adopt $\mu = 2 \pi T$ as a typical renormalization scale. 
 To see the quantitative prediction of the Pad\'{e} summation, 
 we plot, in Fig.3, four different PA's for $R$ as a function of $\alpha_s$
 with $\mu = 2 \pi T$. All four curves turn out to have
 similar trend:
 
First of all,  they do not deviate more than 10 \% from 
 the  SB limit even at $T=250 $ MeV.  This is in 
 contrast to  60 \% deviation of $R$ in eq.(\ref{RR})
 (see Fig.1(a)).
  One should however note that such a small deviation
  does not necessary imply that
  the system can be approximated by  non-interacting gas of quarks and gluons.
 There could be  still strong interactions, but various effects 
  tend to cancel for free energy.  

 Secondly, PA's for the pressure (= $-$ free energy)
 in Fig. 3 show that  overall attraction (repulsion) occurs 
 at high (low) $T$ by the quark-gluon interactions.

 To compare PA's with the original series,
 we plot the result of [4/2]-CPA by the dashed line 
 in Fig.1(a) as an example. This shows the above two aspects
 explicitly.

 Our resummation is solely based on the perturbative coefficients
 calculated through $O(g^5)$ where the non-perturbative 
 magnetic-mass of $O(g^2 T)$  does not play a role.
  Therefore, prediction of PA's in this paper
 could be substantially modified in  the strong coupling (low $T$)
 regime. Nevertheless, it will be interesting to estimate 
 the $O(\bar{g}^6)$ coefficient by our PA's.
  The formulas become particularly simple
 for the cases [3/2]-CPA and [2/3]-CPA, in which 
  $f_6 = f_4 f_5 / f_3 $ and 
  $f_6 = f_2^3 + f_3^2 + (f_4 - f_2^2)/f_3$ are obtained, respectively.
 In both cases, $f_6$ has a proper structure, namely 
 $a + b \ln \bar{g}$, since only $f_4$ depends linearly on 
 $\ln \bar{g}$. 
 This Pad\'{e} prediction will work only for a 
   part of  $f_6$ which is insensitive to the physics of $O(g^2 T)$.

 One should also note that 
 we have not attempted to study the convergence property of PA's.
   This is because low-order PA's do not
 `postdict' the known perturbative coefficients and thus
 do not capture the essential features of the full result
 as we mentioned.
  
In summary,
 we have examined Pad\'{e} approximants of the 
 free energy of hot QCD.  Naive PA's as well as constrained PA's
  reduce the renormalization-scale dependence 
 substantially even at temperature as low as 250 MeV.
 After the Pad\'{e} summation, the free energy takes a value close to the
 SB limit even at 250 MeV. Similar analyses with 
 finite chemical potential and finite quark masses, which 
 are more relevant to the data in relativistic heavy-ion
 collisions,  remain as future  problems \cite{future}.

 After the submission of this work,
 the author was informed about a recent paper treating
   NPA's in QCD and $\phi^4$ theory (B. Kastening, hep-ph/9708219).
  CPA's and EPA's are not studied in this paper.

\section*{Acknowledgements}

 The author thanks M. Gyulassy for 
 his talk at the international workshop on Quark-Gluon Plasma (Kyoto, 
 June 9-11, 1997)
 which leads me to this work. He also thanks 
 T. Kunihiro for helpful discussions.
 This work was supported by
  the Grants-in-Aid of the Japanese Ministry of 
Education, Science and Culture (No. 06102004).

\newpage

\begin{description}
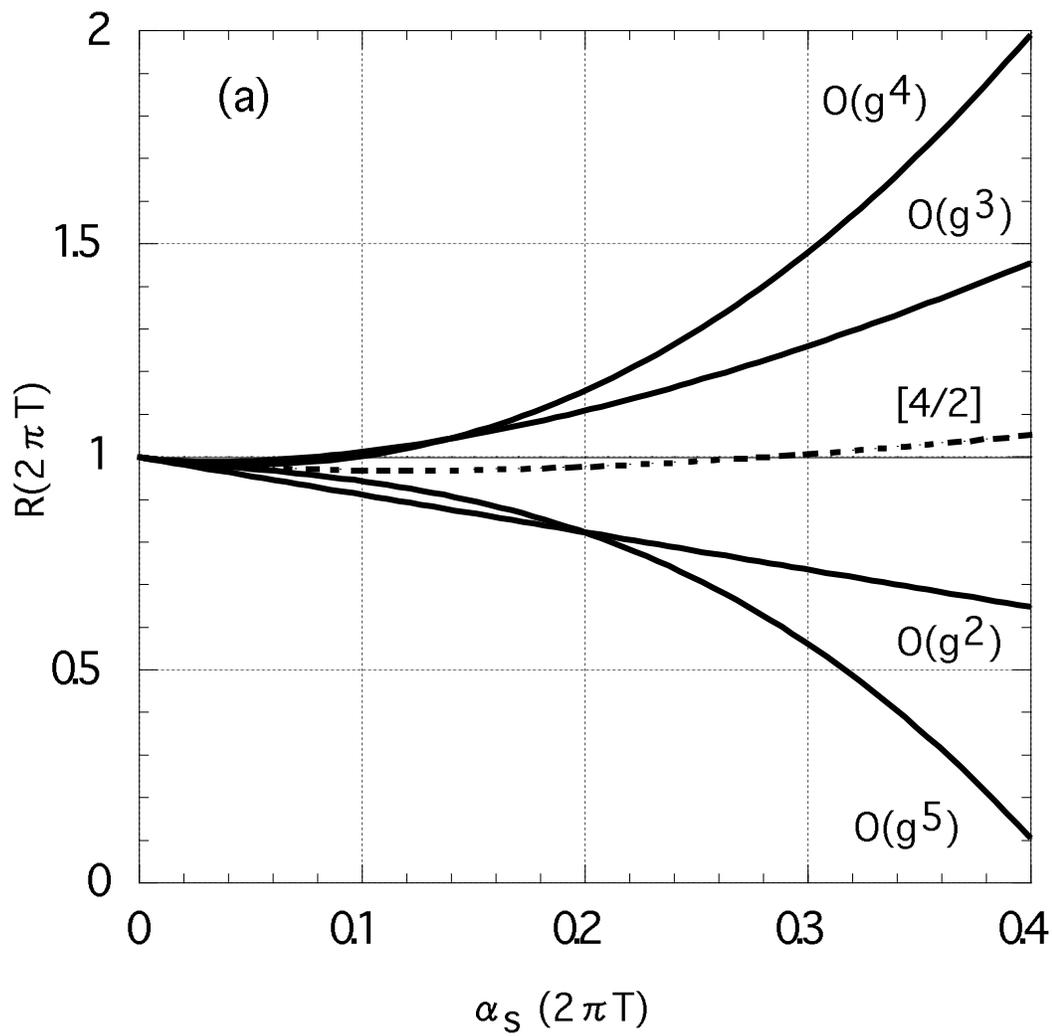
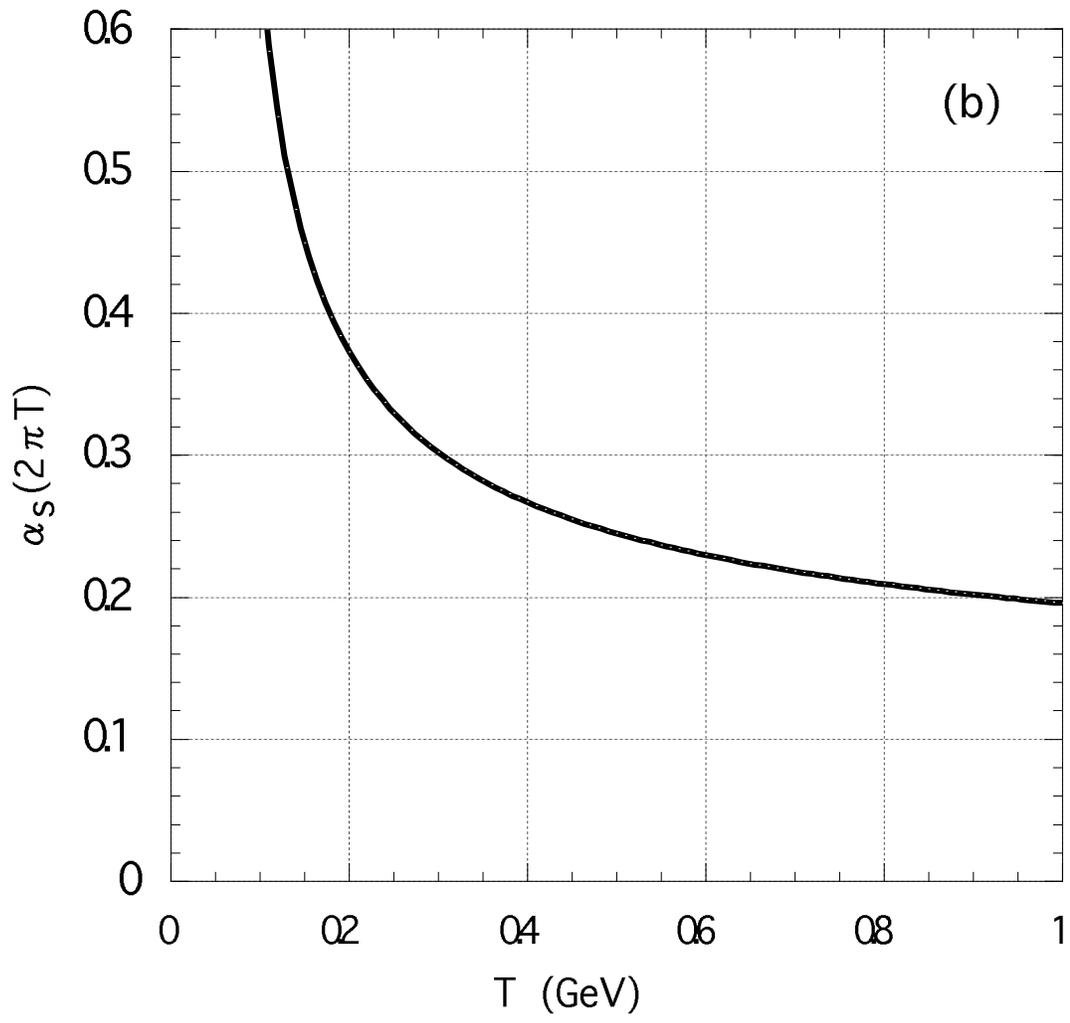
\item[Fig.1:]
 (a) Normalized free energy $R(2 \pi T)$ as 
 a function of the running coupling $\alpha_s(2 \pi T)$.
 $O(g^l)$ ($l=2,3,4,5$) shows  $R$ summed up to
 the $l$-th order.  The dashed line is the result of [4/2] 
 Pad\'{e} approximants.  (b) $\alpha_s(2 \pi T)$ as a function
 of $T$ numerically calculated with the two-loop $\beta$-function.

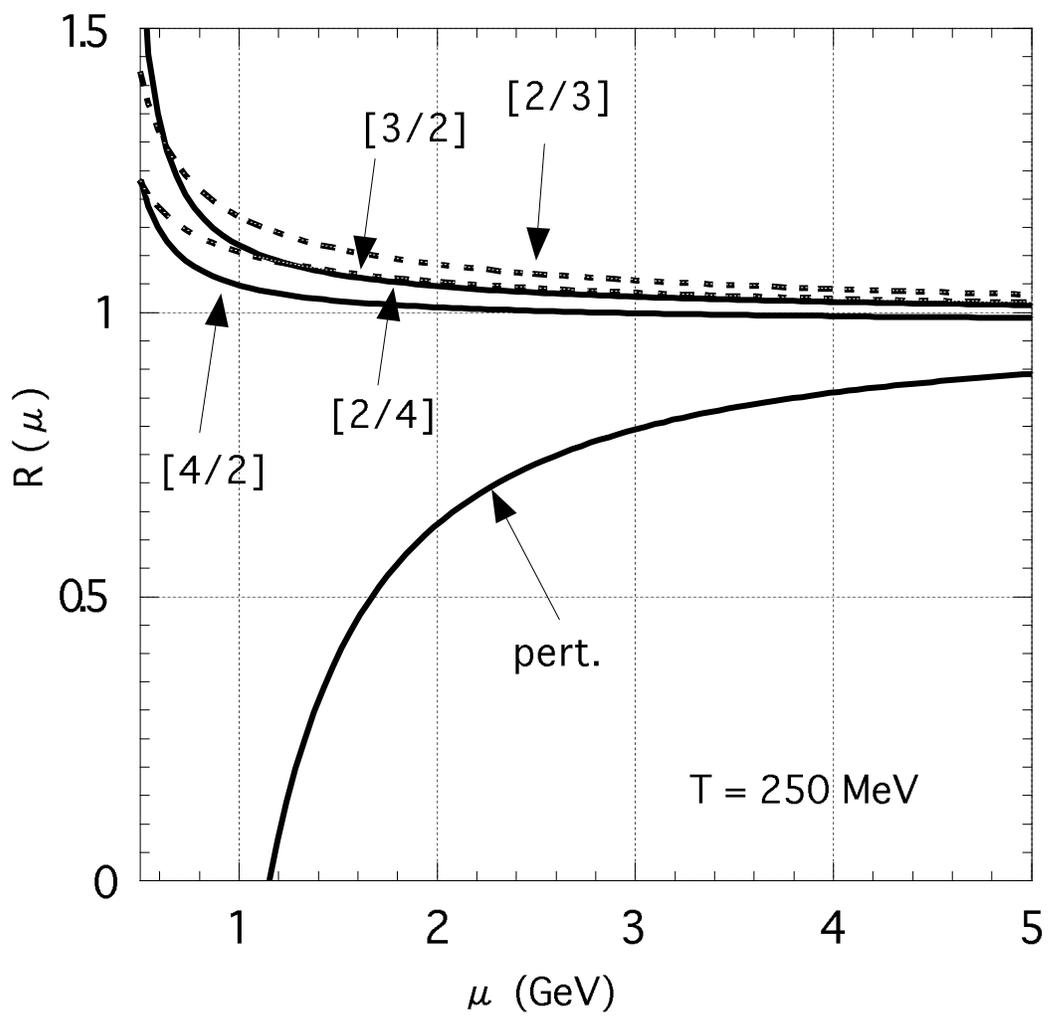
\item[Fig.2:]
 Renormalization-scale dependence of the
 normalized free energy $R(\mu)$. The solid line with
 `pert.' corresponds to the perturbative evaluation
 up to $O(g^5)$. Others 
 are the results after Pad\'{e} summation. 

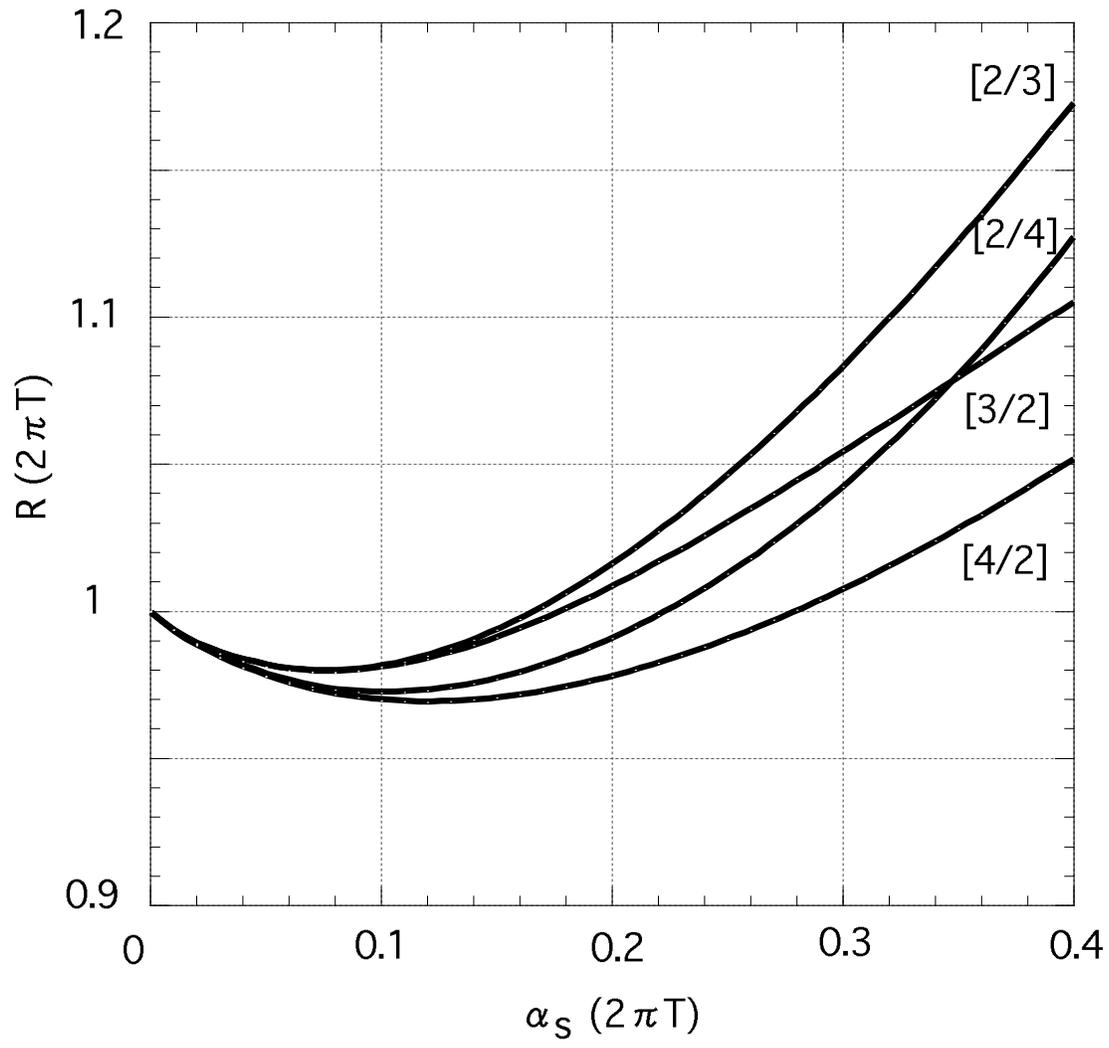
\item[Fig.3:] Pad\'{e} approximants of  $R(2 \pi T)$ as 
 a function of $\alpha_s(2 \pi T)$.

\end{description}


\begin{references}


\btem{review} Quark Matter '96, Nucl. Phys. {\bf A610} (1996).

\btem{asy} J. C. Collins and M. J. Perry, Phys. Rev. Lett. {\bf 34},
 1353 (1975). 

\btem{linde} A. D. Linde, Phys. Lett. {\bf 96B}, 289 (1980);
 D. J. Gross, R. D. Pisarski and L. G. Yaffe, Rev. Mod. Phys.
 {\bf 53}, 43 (1981).

\btem{g5} P. Arnold and C. Zhai, Phys. Rev. {\bf D51}, 1906 (1995).
 C. Zhai and B. Kastening, Phys. Rev. {\bf D52}, 7232 (1995).

\btem{BN} E. Braaten and A. Nieto, Phys. Rev. {\bf D53}, 3421 (1996).

\btem{ah}
 K. Kanaya, Nucl. Phys. B (Proc. Suppl.) {\bf 47}, 144 (1996).
For energy density and pressure above $T_c$ measured on the lattice,
 one needs some careful treatment to extract sensible numbers, see 
 M. Asakawa and T. Hatsuda, Phys. Rev. {\bf D55}, 4488 (1997). 
 
\btem{pade} G. A. Baker and P. Graves-Morris, 
 {\em Pad\'{e} Approximants, 2nd edition},
 (Cambridge Univ. Press, Cambridge, 1996).
  G. A. Baker,
 {\em Quantitative Theory of Critical Phenomena}, (Academic Press,
 New York, 1990).

\btem{ek} J. Ellis, E. Gardi, M. Karliner and M. A. Samuel,
 Phys. Rev. {\bf D54}, 6986 (1996);
 S. Brodsky, J. Ellis, E. Gardi, M. Karliner and M. A. Samuel,
 hep-ph/9706467 and references therein.
 See also, A. L. Kataev and V. V. Starshenko, Mod. Phys. Lett.
 {\bf A10}, 235 (1995).

\btem{gardi} E. Gardi, Phys. Rev. {\bf 56}, 68 (1997). 


\btem{PDG} Review of Particle Properties, Phys. Rev. {\bf D54}, (1996).

\btem{PMS}  P. M. Stevenson, Phys. Rev. {\bf D23}, 2916 (1981).

\btem{BLM} S. Brodsky, G. P. Lepage and P. Mackenzie, Phys. Rev. {\bf D28},
 228 (1983).

\btem{future} The Pad\'{e} approximants based on the perturbation
 series through $O(g^4)$ at finite chemical potential (with zero temperature) 
 as well as a generalization of the analyses in the present 
 paper will be given in a separate publication: T. Hatsuda
  in preparation.
 


\end{references}
\end{document}